\documentclass[12pt]{article}
\usepackage{a4,epsfig,psfig,wrapfig,amsmath,amstext,pstricks}

\begin{document}

\title{
Prospective Analysis of Spin- and CP-sensitive Variables in \mbox{H $\rightarrow$ ZZ $\rightarrow$ l$^+_1$l$^-_1$l$^+_2$l$^-_2$} with \sc{Atlas}
}

\author{C. P. Buszello\footnote{buszello@cern.ch}, I. Fleck\footnote{Ivor.Fleck@cern.ch}, P. Marquard\footnote{marquard@physik.uni-freiburg.de}, J. J. van der Bij\footnote{jochum@phyv1.physik.uni-freiburg.de}\\[0.5cm]
\it Albert-Ludwigs-Universit\"at \\
\it Fakult\"at f\"ur Mathematik und Physik,
\it Physikalisches Institut,\\
\it Hermann-Herder-Stra\ss e 3, D-79104 Freiburg Germany\vspace{0.6cm}\\
}

\date{May 13, 2003}
\maketitle

\vspace{-10cm} \hfill Freiburg-THEP 02/16
\vspace{10cm}
\renewcommand{\arraystretch}{1.2}
\begin{abstract}
A possibility to prove spin and CP-eigenvalue of a Standard Model (SM) Higgs boson is 
presented. We exploit angular correlations in the subsequent decay H $\rightarrow$ ZZ 
$\rightarrow$ 4l (muons or electrons) for Higgs masses above 200\,GeV.
We compare the angular distributions of the leptons originating from the SM Higgs
with those resulting from decays of hypothetical particles with differing quantum numbers.
We restrict our analysis to the use of the Atlas-detector which is one of two multi-purpose 
detectors at the upcoming 14\,TeV proton-proton-collider (LHC) at CERN.  
By applying a fast simulation of the Atlas detector it can be shown that these
correlations will be measured sufficiently well that consistency 
with the spin-CP hypothesis 0+ of the Standard Model can be verified and the 0- and 
1$\pm$ can be ruled out with an integrated luminosity of 100 fb$^{-1}$. 

\end{abstract}
\pagebreak

\section{Introduction}
 
Although the standard $SU(2)_L \times U(1)_Y$ electroweak gauge theory
successfully explains all current electroweak data, the mechanism of spontaneous
symmetry breaking has been tested only partially.
Since in the Standard Model spontaneous symmetry
breaking is due to the Higgs Mechanism, the search for a Higgs particle
will be one of the main tasks of future colliders.\\
 
At present, LEP gives a lower limit of 114.4 GeV/c$^2$ \cite{lephiggs}
for the Higgs boson mass. There is also an indirect upper limit from electroweak 
precision measurements of 219 GeV/c$^2$ at a 95\% confidence level \cite{zeuthen}
which is valid in the minimal standard model.
However, this limit is still preliminary and the quality of the SM fit,
when including all EW measurements from both low and high energy
experiments, is still an object of discussion amongst experts \cite{moenig}. 
Furthermore, 
a heavier Higgs boson would be consistent with the electroweak precision
measurements in models more general than the minimal standard model \cite{peskin}.
In this first analysis we will therefore also consider  higher 
Higgs masses well above this limit, as
can be produced at high energy hadron colliders such as the LHC (pp
collisions at 14 TeV).
A Standard Model  
Higgs boson lying below the $WW$ threshold will mainly decay into
a $b\bar b$ pair. In this case, there is
an overwhelming direct QCD background which dominates the signal.
Therefore, the Higgs boson is difficult
to study in detail in this mass region, even though one can use rare decays
as a signal.
Rare decays considered in the literature include,
for example, $H \to \tau^{+}\tau^{-}$, $H \to \gamma\gamma$,
$H \to Z\gamma$, $H \to ZZ^{*}$ 
or $H \to WW^*$.
All of these signals are rather difficult to see, but can eventually
be used to establish the existence of the Higgs boson \cite{piccinini}.
While these decay modes can be used to discover the Higgs boson,
a detailed study of its properties will be difficult.\\


The situation is much better for a heavy Higgs boson ($m_H>2m_W$).
For such a Higgs boson the main decay products are vector boson pairs,
$W^{+}W^{-}$ or $ZZ$.
For the latter decay mode, a clear signal for the Higgs 
consists of a peak in the invariant mass spectrum of the
produced vector bosons.
The double leptonic decay of the $Z$ boson, $H \to ZZ\to l^{+}_{1}l^{-}_{1}l^{+}_{2}l^{-}_{2}$,
leads to a particularly clean signal.\\
 
In this case, the basic strategy
for discovering a Higgs boson in a clean mode is to select 
events with 
4 high P$_T$ leptons that can be combined to form
two Z-bosons. Here, an exposure of 30 fb$^{-1}$ is already sufficient.
If one finds such a signal 
one might be tempted to assume this to be the
Standard Model Higgs boson. However, given the fact that the 
Higgs sector is not fully prescribed, one
has to allow for other possibilities. In
strongly interacting models, for instance, low lying (pseudo-)vector resonances
are possible\cite{dominici,kastening}. Also, pseudoscalar particles
 are present
in a variety of models \cite{higgsreview}. Therefore, the first priority
after finding a signal is to establish the
nature of the resonance, in particular its spin
and CP-eigenvalue. This can be done by studying angular
distributions and correlations among the decay leptons.
In the following, we will make this study. We will limit
ourselves to (pseudo-)vector and (pseudo-)scalar particles.
To demonstrate consistency with a Spin 0, CP even hypothesis,
we will compare the angular distributions, to those 
produced by different particles, always assuming the 
production rate of a Standard Model Higgs boson. 
This is the right assumption to make because,
in order to be recognized as a candidate for a 
Standard Model like Higgs,
the detected signal must be a resonance with
the appropriate width and branching ratios.
Since the production mechanism - 
gluon-fusion rules out spin 1 particles, due to Yang's theorem\cite{yang} -   
cannot be seen, the only way to prove that the spin and CP nature of the new particle 
is Standard Model like is to study the decay angles of the leptons.\\


Theoretical studies of angular distributions have been performed in the 
literature [9-15].
So far, such studies have been limited
to theoretical discussions. However, it was shown
in \cite{bij} that acceptance and efficiencies
of the detector can play a role since they can generate
correlations, mimicking physical ones. Therefore, it is necessary
to use a detector simulation in order to establish how well
one can do in practice.\\

The complete triple differential cross-sections for a Higgs-boson decaying 
into two onshell Z-bosons which subsequently decay into fermion 
pairs can be calculated at tree level. The angular dependence of this 
cross section is given in the appendix together with the most important  
integrated angular distributions. 
For the definition of the angles see Figure 1.\\


\begin{figure}
 \epsfig{figure=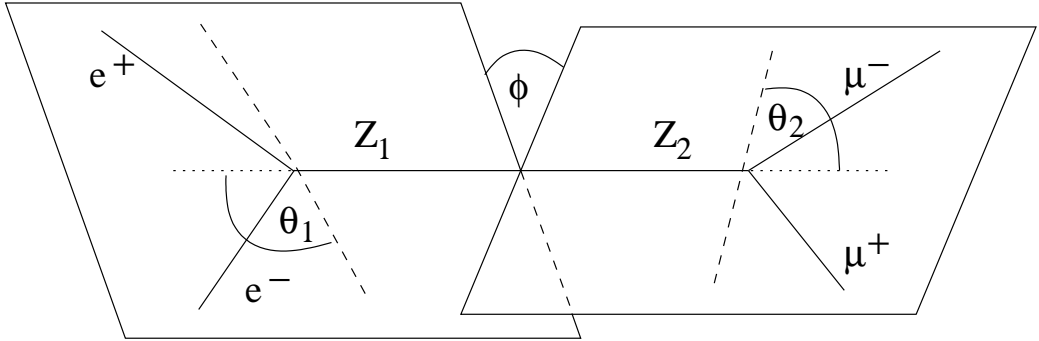,width=\textwidth}
 \caption{The decay plane angle $\phi$ is measured between the two planes defined by the leptons 
 from the decay of the two Z
 bosons in the  rest frame of the Higgs, using the charge of the leptons to fix the orientation 
 of the planes. 
 The dashed lines represent the direction of motion of the 
 leptons in the rest frame of the Z Boson from which they originate. The angles $\theta_1$ and $\theta_2$ 
 are measured between the negatively charged leptons and the direction of motion of the 
 corresponding Z in the Higgs boson rest frame. $\phi$=0 correspond to $p_{e^+}\times p_{e^-}$
 and  $p_{\mu^+}\times p_{\mu^-}$ being parallel. $\phi$=$\pi$ correspond to $p_{e^+}\times p_{e^-}$
 and  $p_{\mu^+}\times p_{\mu^-}$ being antiparallel.}
 \label{figplanes}	
 \end{figure}

We study essentially two distributions.
One is the distribution of the cosine of the polar angle,  $\cos\theta$,
of the decay leptons relative to the $Z$ boson.
Because the heavy Higgs decays mainly into longitudinally
polarised vector bosons the cross section $d\sigma / d\cos\theta$
should show a maximum around $\cos(\theta$)=0.
The other is the distribution of the angle $\phi$ between the decay
planes of the two $Z$ bosons in the rest frame of the Higgs boson.
This distribution depends on the
details of the Higgs decay mechanism.
Within the Standard Model, a behaviour roughly like
$1 + \beta~\cos 2\phi$ is expected.
This last distribution is flattened in
the decay chain $H \rightarrow ZZ \rightarrow 4l$, because of the small
 vector coupling of the leptons, in contrast to the decay of the Higgs Boson into W's or
 decay of the Z into quarks. Also, cuts can significantly affect the correlations.
Therefore one needs a precise measurement of the momenta of the outgoing leptons.
 The Atlas-detector should be 
well-suited to measure these distributions, since the muon and electron reconstruction 
is very precise over a large solid angle. 
A detector Monte-Carlo is however needed in order
to determine whether the angular distributions can be 
measured sufficiently well in order to determine the quantum numbers of the Higgs particle.\\

The present paper is organized as follows. In Chapter 2 the generator is described,
in Chapter 3 detector simulation and reconstruction are given.
In Chapter 4 we define quantities that can be used to characterize the different 
distributions. In Chapter 5 we present the results, concluding that the
quantum numbers of the Higgs particle can indeed be determined. 
In the appendix
we give formulae for the complete differential and integrated distributions
for the decay of the resonance assuming arbitrary couplings computed in tree level and narrow 
width approximation.

\section{The Generators}

In order to distinguish between different \mbox{spins J=0,1} and/or CP-eigenvalues 
$\gamma_{CP}=-1,+1$ one needs to study four different distributions: 
that resulting from the decay of the Standard Model Higgs boson, and the three distributions that would result from hypothetical particles with spin and CP-eigenvalue combinations
(0, 1), (1, 1), (1, -1).


The feasibility of using angular correlations in the decay of the Z bosons   
in order to distinguish between these particles
has been evaluated  using two different Monte-Carlo ge\-ne\-ra\-tors. 
One was written for the Standard Model Higgs 
($gg \rightarrow H \rightarrow ZZ \rightarrow 4l$) and the irreducible ZZ-background.\cite{bij} 
The latter includes contributions from both gluons and quarks to the ZZ production
( \mbox{$gg \rightarrow ZZ \rightarrow 4l$} and \mbox{$q\bar{q} \rightarrow ZZ \rightarrow 4l$})
, whereas all Higgs production mechanisms other than the gluon fusion are neglected. 
This generator keeps all polarisations of  the Z-boson
for the quark initiated as well as for the gluon initiated processes. 
This allows for
an analysis of the angular distributions of the leptons. The  gluonic production of Z-boson pairs is only about 30\% of the total background. 
However, one should not ignore its contribution, since it has different angular distributions from
the other backgrounds and its presence can affect measured correlations.
The programme contains no K-factors, therefore our conclusions regarding the feasibility of the determination
of Higgs quantum numbers are conservative.
 Indeed, K-factors are expected to be larger
for gluon-induced processes (such as the signal) than for quark-induced
processes (70\% of the background).
Since the narrow width approximation
is used, the results are only valid for Higgs masses above the ZZ threshold.\\

For the alternative particles, a new generator was written
based on an article by C. A. Nelson and J. R. Dell' Aquilla \cite{distri4}.
The programmes for the production of background, the Standard Model Higgs and
all alternative particles use
Cteq4M  structure functions \cite{cteq} and hdecay \cite{spira} for 
branching-ratio and width of the Higgs, and all use the narrow width approximation.
The background as well as all cross sections for the four 
simulations are taken from the first generator. Thus, all cases show identical 
distributions of invariant mass of the Z-pairs and transverse momentum $P_{T}$ of Z-bosons and 
leptons and have the same width  and cross sections. The only difference lies in  the angular 
distributions of the leptons. For the alternative particles no special assumption concerning 
the  coupling has to be made; only CP-invariance is assumed. It is worth mentioning 
that the angular correlations are completely independent of the production mechanism.\\

\section{Detector Simulation and Reconstruction}

The detector response is simulated using ATLFast \cite{atlfast}, a software-package for particle 
level simulation of the Atlas detector. It is used for fast event-simulation including the most crucial 
detector aspects. Starting from a list of particles in the event, it provides a list of 
reconstructed jets, isolated leptons and photons and expected missing transverse energy. 
It applies momentum- and energy- smearing to all reconstructed particles.
The values of the detector-dependent parameters are chosen to match the expected
performance that was evaluated mostly by full simulations using Geant3\cite{TDRV1}.\\

The event selection is modeled exactly after the event selection in the Atlas-Physics-TDR 
\cite{TDRV2}. Four leptons (electrons or muons) are required in the pseudorapidity range 
\mbox{$|\eta| = |\ln \tan(\frac{\theta_{Beam}}{2})|\,<\,2.5$} ($\theta_{Beam}$ being the angle to the 
beam axis). Two of the  leptons are required to have transverse momenta greater than 20 GeV/c 
and the two other leptons must have transverse momenta greater than 7 GeV/c each. 
A lepton identification efficiency of 90\% per lepton was assumed. 
Two Z 
bosons are reconstructed by choosing lepton-pairs of matching flavour and opposite sign. 
If the flavours of all four leptons are equal, the combination is chosen, which 
minimizes the sum of the squared deviation of the invariant mass of the pairs with opposite sign from the Z mass (i.e. choose combination ab/cd that minimizes (m$_{a^+b^-}$-m$_Z$)$^2$ + (m$_{c^+d^-}$-m$_Z$)$^2$). 
The reconstructed invariant mass of the two reconstructed Z bosons has to lie within two times
the width of the reconstructed mass peak of the Higgs resonance around the centre of the peak.
For high Higgs masses ($m_H >$  300\,GeV/c$^2$) this is only little more than two times the decay width, while for smaller masses the experimental resolution dominates.\\

Throughout this paper, we use the term signal for distributions where the background 
has been statistically subtracted. 
The only background considered is the Z pair production. 
Other possible backgrounds 
like top pair production or Zbb are negligible for masses of the Higgs boson above 200 GeV/c$^2$.   
Systematic uncertainties due to the simulation of the background could be studied by comparing
distributions from the sidebins of the Higgs signal with the results of the generator.
A proper treatment of the background is very important, since the angular distributions 
of the background itself and correlations introduced by detector effects have a large impact 
on the shape of the distributions discussed. These effects are detailed below.\\

\begin{figure}
\epsfig{figure=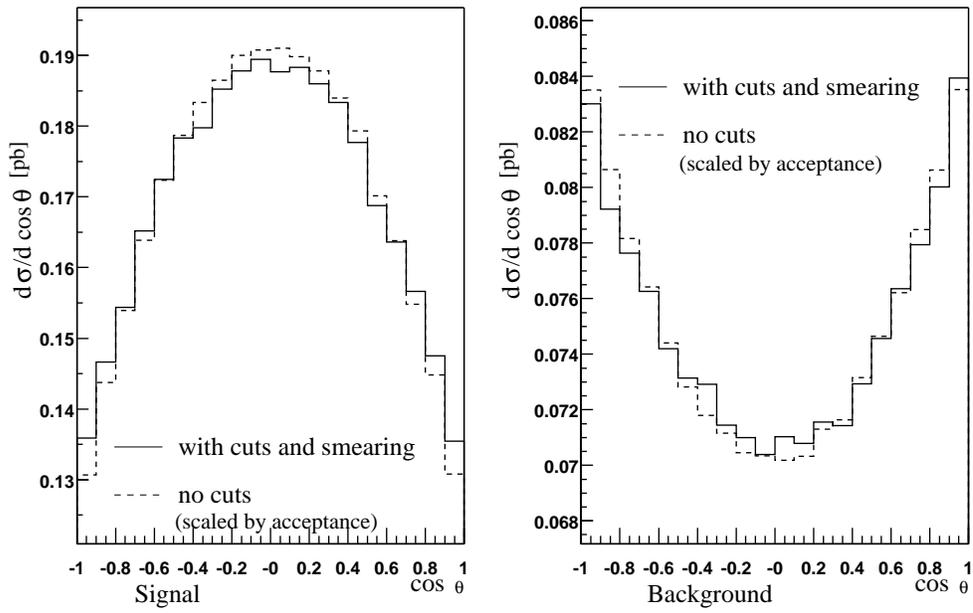,width=\textwidth}    
\caption{Distribution of the polar angle $\cos(\theta)$ for the background only (right) and the signal (left). The Higgs mass is 200 GeV/c$^2$.}
\label{figpolar1}
\end{figure}

\begin{figure}[h]	
\epsfig{figure=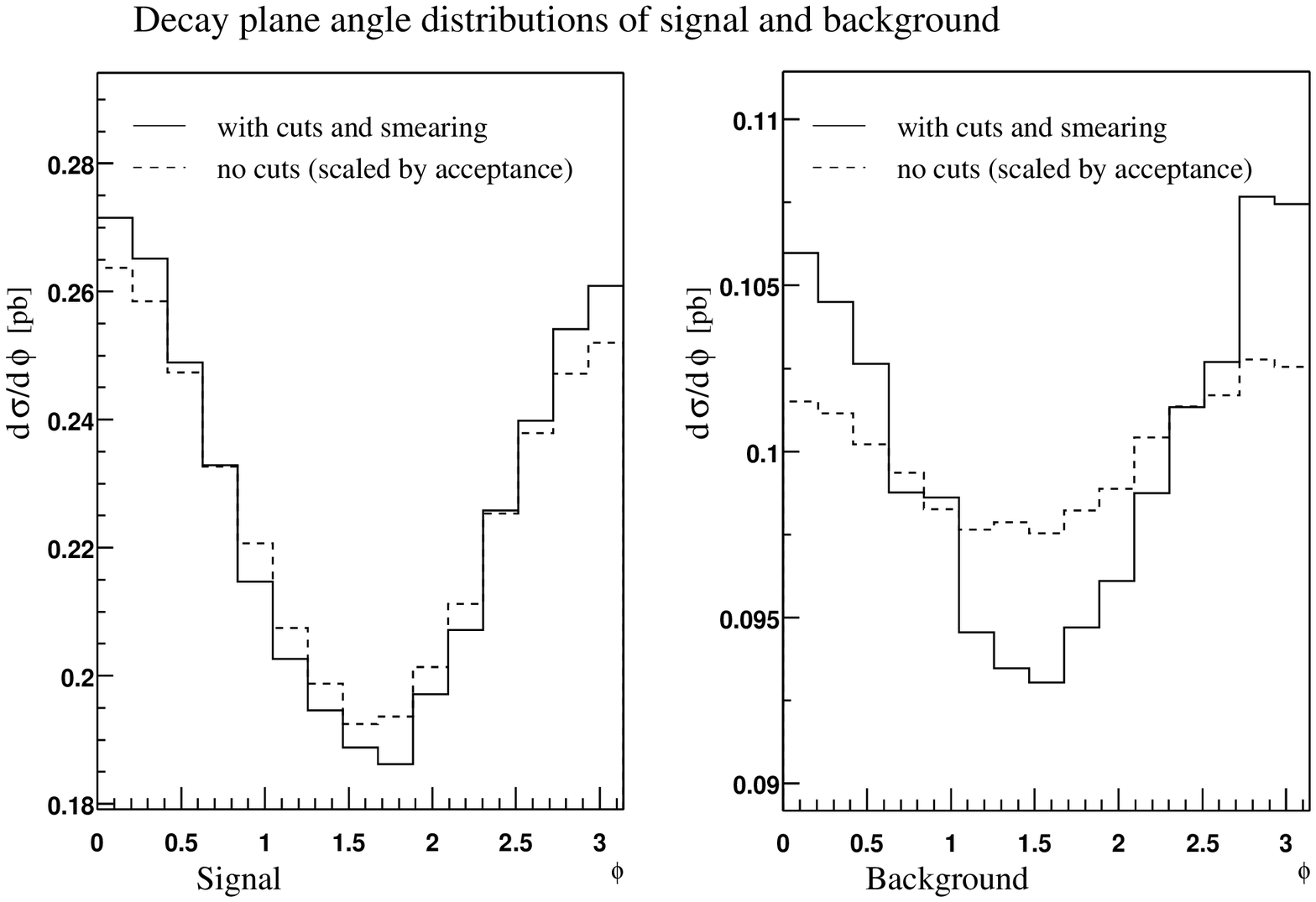,width=\textwidth}
\caption{Distribution of the decay plane angle $\phi$ for the background only (right) and the 
signal (left). The Higgs mass is 200 GeV/c$^2$. The distribution without cuts is scaled by the expected acceptance, so that the shape can be directly compared. }
\label{figplane1}	
\end{figure}

For high invariant masses,
the Z bosons from the background processes are mainly transversely polarised leading to a polar angle 
distribution  of the form $\frac{d\sigma}{d\cos\theta} \sim 1 + \cos^2 \theta $. This distribution flattens 
the $\sin^2 \theta $ distribution expected for the Higgs decay.
Figure \ref{figpolar1} shows the polar angle distributions of the signal (left) and the background (right).
The dashed line shows the shape of the distribution expected when no cuts are applied and the detector response is not taken into account. 
It has just been scaled by the overall acceptance of the cuts, so that the shape can be compared.
The expected distribution with all cuts and smearing applied is drawn as a solid line. 
Figures \ref{figpolar1} and \ref{figplane1} are produced assuming a 
Higgs mass of 200\,GeV/c$^2$ and Z decaying to muons only. 
For the decay ZZ $\rightarrow$ e$^+$e$^-$e$^+$e$^-$ or ZZ $\rightarrow$ $\mu^+\mu^-$e$^+$e$^-$  the graphs look similar. The smearing effects are largely
independent of the Higgs mass.\\

The effect of the detector acceptance and isolation cuts on the decay plane angle 
distribution is shown in Figure \ref{figplane1}.
Again, the distribution of the signal is shown left and 
the background right.
The dashed histogrammes are scaled to have the same integral
as the solid histogrammes, and zero is suppresssed in 
order to facilitate the 
comparison of the shape. The definition of the line styles
are the same as above.
For the decay plane angle, the background shows an almost flat distribution before applying
selection cuts. But a minor correlation
is introduced by detector effects. This has been  simulated and taken into account for the 
analysis of decay plane angle distributions. 
In conclusion, the isolation cuts lead to a small distortion of the 
angular distributions as discussed in 
\cite{bij}, but these effects are almost negligible for the Atlas detector. The cut on $|\eta|$ enhances 
the decay plane correlation a little, but the smearing and the $P_{T}$ requirements 
reduce this effect. Altogether, there is a small enhancement of the correlation of almost the same 
amount for all four particles.\\

A further cut on the transverse momentum of the Z bosons $P_{T}^{max}(Z_1,Z_2) > m_H\,/\,3$ is 
known to additionally reduce the 
background, but it also affects the correlation. Since an optimisation of the signal-to-background 
ratio is not crucial to this analysis, this cut has not been applied, rendering the analysis
less dependent on the details of the production mechanism like initial $P_T$ of the Higgs boson.\\

\section{Parametrisation of Decay Angle Distributions}
\label{chap:para}
The differential cross sections for the different models can be computed directly or can be 
derived from the formulae given in \cite{distri4}. The explicit distributions are given in the appendix.
From the article \cite{distri4} we quote the simple distributions 
of the alternative particles.  
Table \ref{distriint1} shows the distribution of the polar angle $\theta$.
$\theta_1$ and $\theta_2$ are the polar angles of the leptons originating from 
the Z Bosons $Z_1$ and $Z_2$ respectively.  
In Table \ref{distriint2},  the distribution of the decay plane angle $\phi$ is shown
where the polar angle $\theta$ is integrated over different ranges.
F11 gives the distribution for $ 0 \le \theta_{1,2} \le \pi/2 $, 
F22 for $ \pi/2 \le \theta_{1,2} \le \pi $, 
F12 for $ 0 \le \theta_{1} \le \pi/2 $ and $ \pi/2 \le \theta_{2} \le \pi$,
and F21 for $ \pi/2 \le \theta_{1} \le \pi $ and $ 0 \le \theta_{2} \le \pi/2 $.
$ \cal R$, $\cal U$, $\cal T$ and $ \cal W $ are the parameters that characterise the decay density matrix.
For the decay modes used in this analysis, they amount to the following values:
$\cal R$ = $\cal U$ = -1/2, $\cal T$ = $ \cal W $ = $- \frac{2 r }{1 + r^2}$. r is the ratio of the axial
vector to vector coupling which for the muons amounts to r = (1\,-\,4\,$\sin^2 \theta_W)^{-1}$. We used $\sin^2 \theta_W$ = 0.23.\\

\begin{table}
\begin{tabular}{|l|l|l|}
\hline
Spin & $\gamma_{CP}$ & \\ \hline \hline
0 & -1 & 1 -  $\cal R$$P_2(\cos \theta_1)$ - $\cal U$$P_2(\cos \theta_2)$ + $\cal R\cal U$$P_2(\cos \theta_1)P_2(\cos \theta_2)$ + \\ &  & $\frac{9}{4}$$\cal T$$ \cal W $$P_1(\cos \theta_1)P_1(\cos \theta_2)$ \\ \hline
1 & +1 & 1 + $\frac{1}{2}$ $\cal R$$P_2(\cos \theta_1)$ + $\frac{1}{2}$ $\cal U$ $P_2(\cos \theta_2)$ -2 $\cal R\cal U$ $P_2(\cos \theta_1)P_2(\cos \theta_2)$\\ \hline
1 & -1 &  1 + $\frac{1}{2}$ $\cal R$$P_2(\cos \theta_1)$ + $\frac{1}{2}$ $\cal U$ $P_2(\cos \theta_2)$ -2 $\cal R\cal U$ $P_2(\cos \theta_1)P_2(\cos \theta_2)$\\ \hline

\end{tabular}
\caption{Distribution of the polar angle $\theta$. $P_{i}$ are the Legendre Polynomials. See the text for 
definitions.}
\label{distriint1}
\end{table}
 
\begin{table}
\begin{tabular}{|l|l|}
\hline
Spin=0 $\gamma_{CP}$=-1& \\ \hline \hline 
F11 + F22 & $1 + \frac{9}{16} \cal{T}  \cal{W}   - \cal {R}\cal{U}$$ \cos(2 \phi) $ \\ \hline
F12 + F21  & $1 - \frac{9}{16} \cal{T}  \cal{W}   - \cal {R}\cal{U}$$ \cos(2 \phi) $ \\ \hline \hline

Spin=1 $\gamma_{CP}$=+1& \\ \hline \hline 
F11 + F22 & $1 + (-\frac{1}{2}$$\cal R\cal U$$+\frac{1}{2}$$\cal T\cal W$$ (\frac{3\pi}{8})^2) \cos(\phi) $ \\ \hline 

F12 + F21 & $1 + (+\frac{1}{2}$$\cal R\cal U$$+\frac{1}{2}$$\cal T\cal W$$ (\frac{3\pi}{8})^2) \cos(\phi)$ \\ \hline \hline

Spin=1 $\gamma_{CP}$=-1& \\ \hline \hline 
F11 + F22  & $1 + (+\frac{1}{2}$$\cal R\cal U$$-\frac{1}{2}$$\cal T\cal W $$(\frac{3\pi}{8})^2) \cos(\phi) $ \\ \hline 
F12 + F21  & $1 + (-\frac{1}{2}$$\cal R\cal U$$-\frac{1}{2}$$\cal T\cal W $$(\frac{3\pi}{8})^2) \cos(\phi)$ \\ \hline

\end{tabular}
\caption{Distribution of the decay plane angle $\phi$.
F11 gives the distribution 
for $ 0 \le \theta_{1,2} \le \pi/2 $, F22 for $ \pi/2 \le \theta_{1,2} \le \pi $, 
F12 for $ 0 \le \theta_{1} \le \pi/2 $ and $ \pi/2 \le \theta_{2} \le \pi $ 
F21 for $ \pi/2 \le \theta_{1} \le \pi $ and $ 0 \le \theta_{2} \le \pi/2 $. 
$\cal R$ = $\cal U$ = -1/2 , $\cal T$ = $ \cal W $ = $-\frac{2 r }{1 + r^2}$, r = (1\,-\,4\,$\sin^2 \theta_W)^{-1}$.}
\label{distriint2}
\end{table}

\begin{figure}
\epsfig{figure=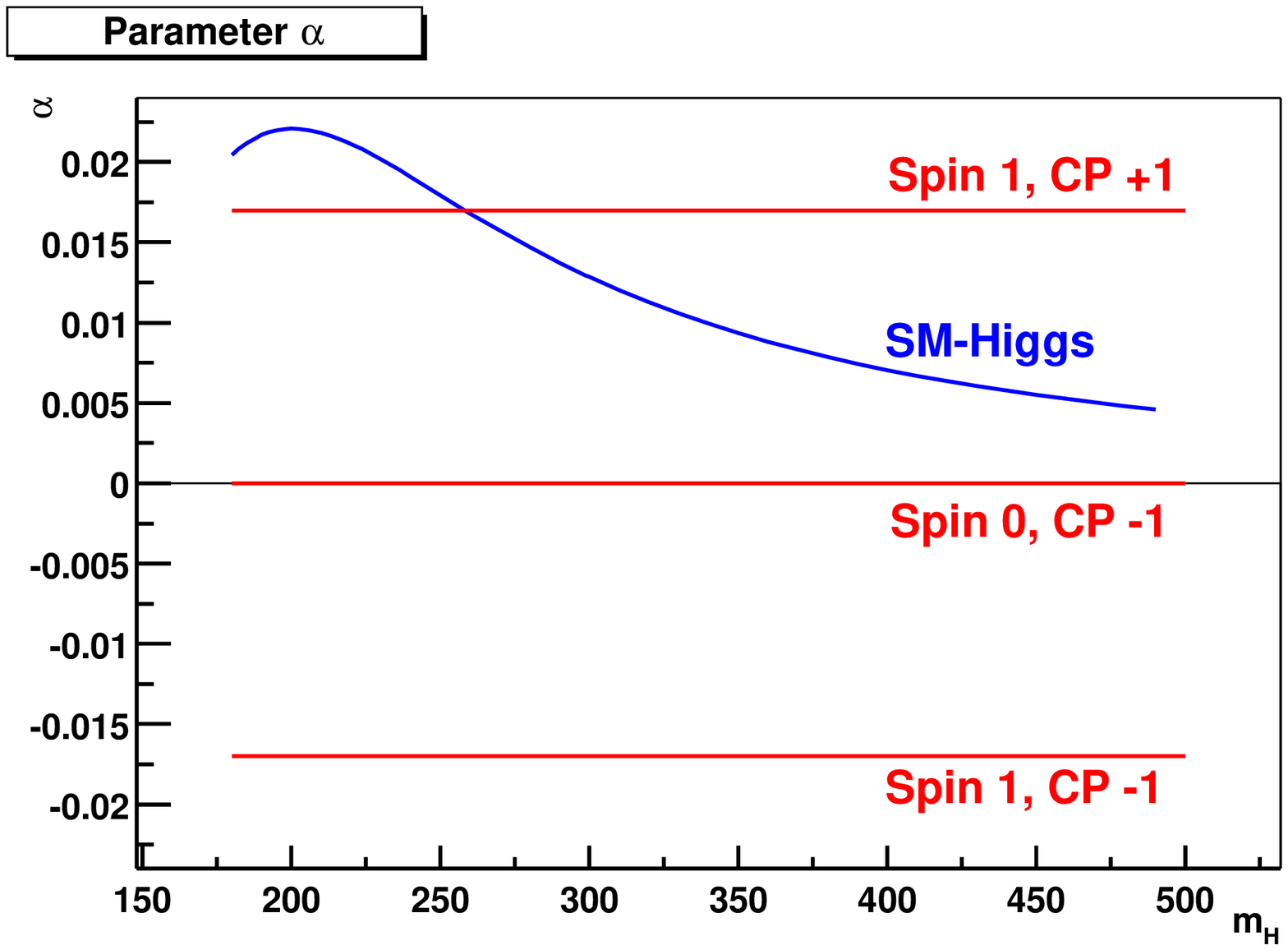,width=0.7\textwidth}
\epsfig{figure=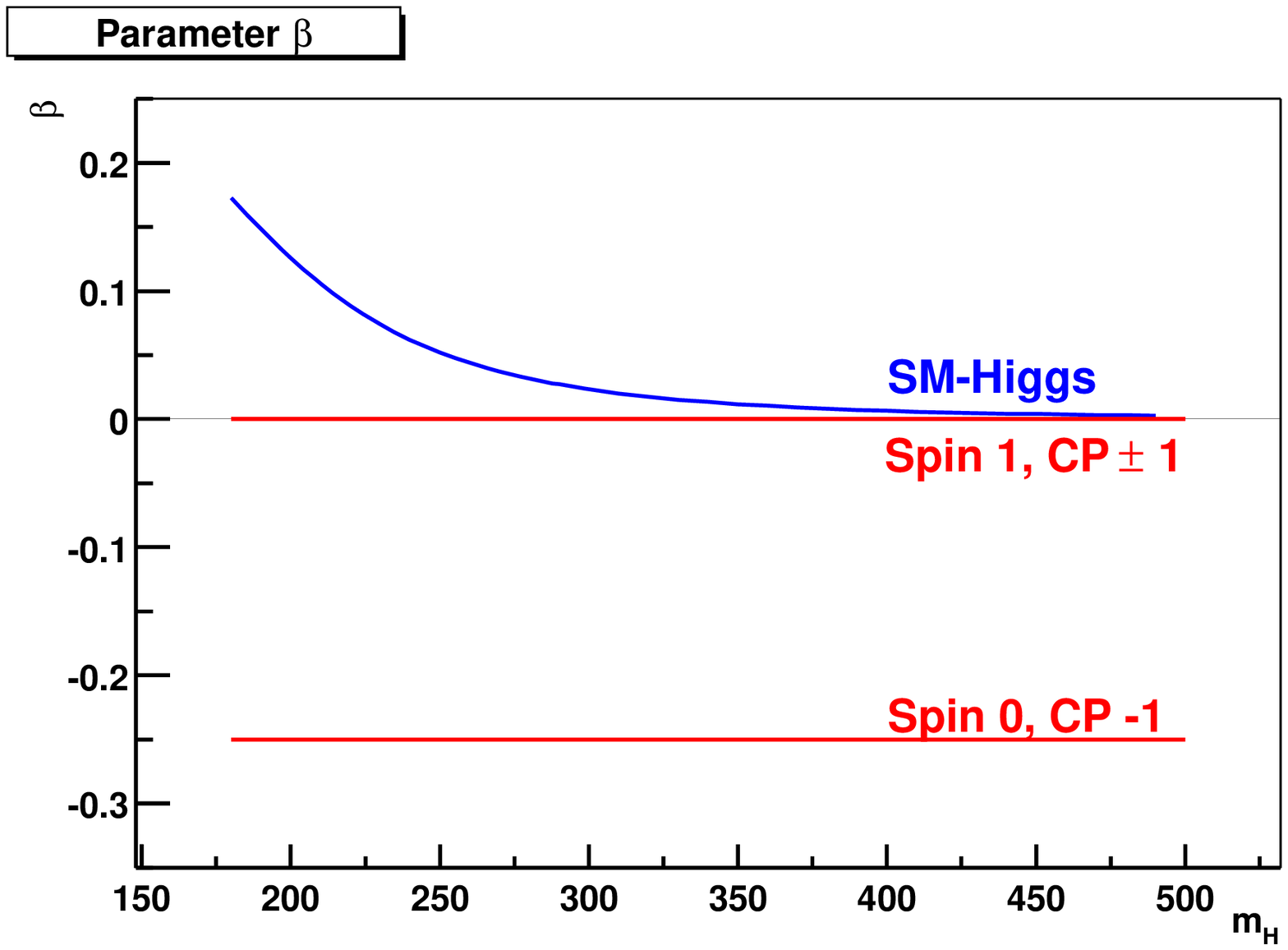,width=0.7\textwidth}
\epsfig{figure=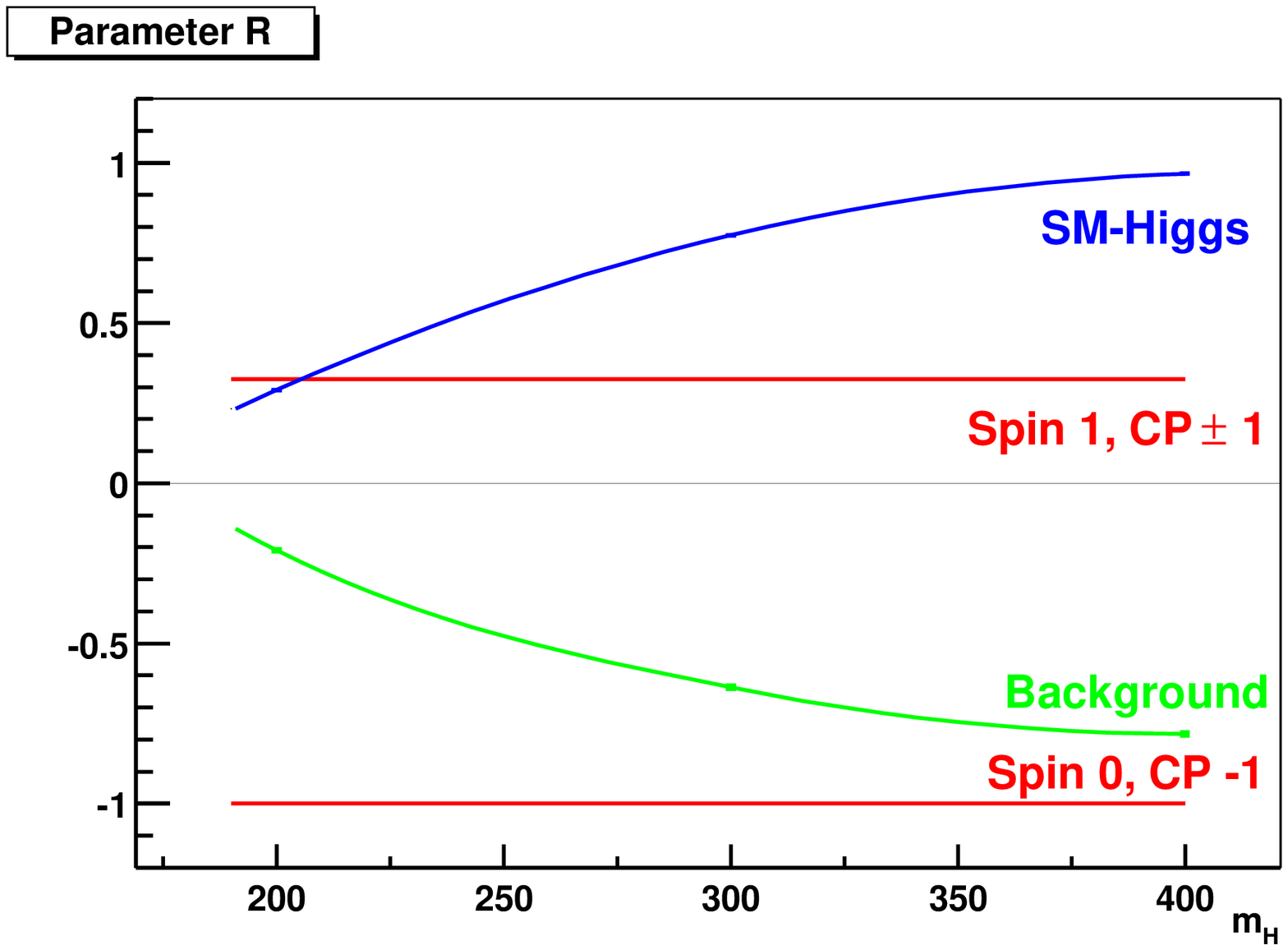,width=0.7\textwidth}
\caption{The variation of the three parameters $\alpha$, $\beta$ and $R$ (top to bottom) with
  the Higgs mass.}
\label{figtheo}
\end{figure}

The plane-correlation can be parametrised as 
\begin{equation}
 F(\phi) = 1 + \alpha \cdot \cos(\phi) + \beta \cdot \cos(2 \phi)
\end{equation}
In all four cases discussed here, there is no $\sin(\phi)$ 
or $\sin(2 \phi)$ contribution. For the Standard Model Higgs, $\alpha$ and $\beta$ depend 
on the Higgs mass  while they are constant over the whole mass range in the other  
cases.\\

The polar angle distribution can be described by 
\begin{equation}
G(\theta) = T \cdot (1 + \cos^2(\theta)) + L \cdot \sin^2(\theta) 
\end{equation} reflecting the longitudinal or transverse 
polarisations of the Z boson. We define the ratio 
\begin{equation}
R := \frac{L - T}{L + T} 
\label{eq:Rdef}
\end{equation}
of 
transversal and  longitudinal polarisation. \\
The dependence of the parameters $\alpha$, $\beta$ and $R$ on the Higgs mass is shown in Figure \ref{figtheo}. 
The pseudoscalar shows the largest deviation from the SM Higgs. It would have 
\mbox{$\beta = -0.25$} and \mbox{$R=-1$} whereas the scalar always has $\beta>0$ and $R>0$. 
The vector and the axialvector can be excluded through the parameter $R$ 
for most of the mass range, but for
Higgs masses around 200 GeV/c$^2$ the main difference lies in the value of 
$\beta$ which is zero for $J=1$ and $\gamma_{CP} = \pm 1$ and about 0.1 for the 
scalar. The value for $\alpha$ can only discriminate between the scalar and the axialvector but
the difference is very small. 


\section{Background Estimation}

\begin{figure}
\epsfig{figure=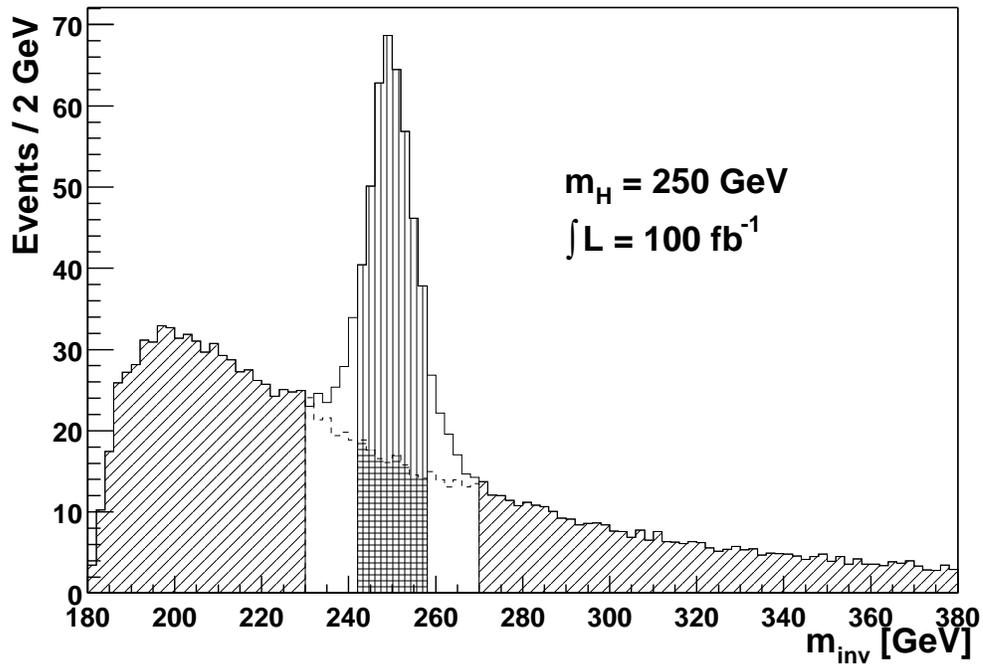,width=\textwidth}
\caption{The invariant mass distribution of a 250 GeV/c$^2$ Higgs boson and the ZZ Background. The vertically hatched region is the signal region used in the analysis. The diagonally hatched regions are the sidebands used to determine the 
expected number of background events (hatched horizontally) inside the signal region. The dotted line indicates the shape of the background in the transition
region between the sidebands and signal which is not used at all.}
\label{fig:sidebands}
\end{figure}
\begin{figure}

\epsfig{figure=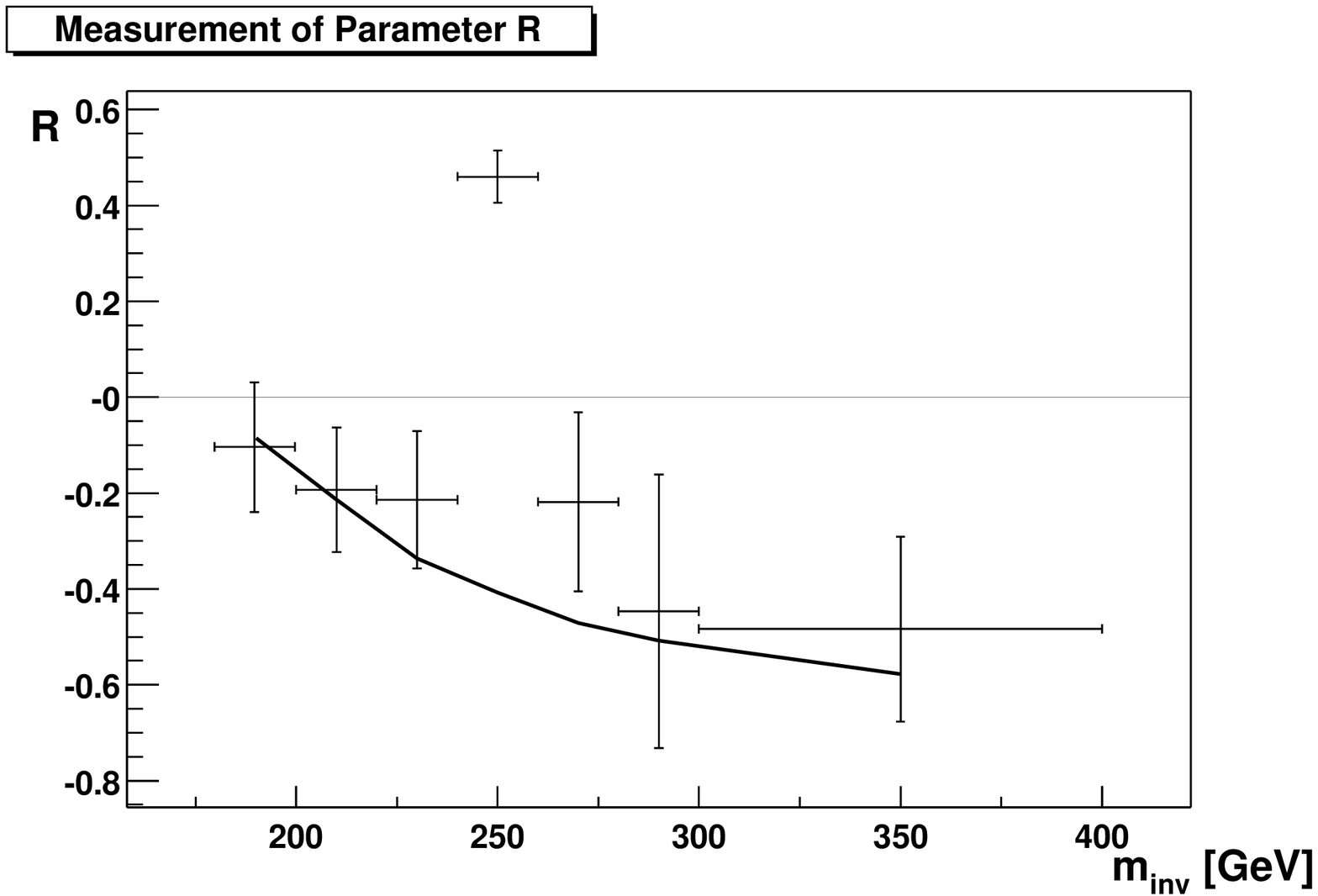,width=\textwidth}
\caption{The parameter $R$ (defined in (3)) as obtained by a fit to different mass regions of 
the background only (solid line) and by a fit to the same mass regions to 
signal plus background distributions (points with errorbars). The horizontal 
errorbars indicate the regions from which the distributions where taken.   
}
\label{fig:ParR}
\end{figure}

The subtraction of the angular distributions of the background 
is necessary to obtain and analize
the angular distributions of the signal alone.
This bears a risk of introducing systematic errors. Thus, the background
distributions as produced by Monte-Carlo-Generators have to be checked 
against the data. In this chapter we will estimate the effects and 
possible systematic errors introduced by the subtraction.\\

First, the absolute number of background events has to be estimated.
This can be done by comparing the sidebands of the signal to a 
simulated distribution of the background only. This procedure is illustrated 
in Figure \ref{fig:sidebands}. In order to obtain the number of expected events
 the number of simulated events in the signal region N$^{MC}_{signal}$ is scaled by the number 
of events in the sidebands N$^{Data}_{side}$ divided by the number of 
simulated events in the 
sidebands N$^{MC}_{side}$. The error from this calculation is $\sigma_{\rm N}$ = 
$\sqrt{N^{Data}_{side}} \cdot \frac{N^{MC}_{signal}}{N^{MC}_{side}}$. 
In the case of a 250\,GeV/c$^2$ Higgs boson the estimated number of background events is N=130 with a systematic error of $\sigma_{\rm N {\rm syst}}$ = 4.1 which is well below the statistical error of 
$\sigma_{\rm N {\rm stat}}$ = 11.4.\\

Checking of the shape of the background distribution can be done 
by using bins below and above the signal region, too. This is 
demonstrated in Figure \ref{fig:ParR}. It shows the parameter $R$ derived 
from a fit described in chapter  \ref{chap:para} to 
the background distribution only (black line) and the background plus signal
(points with errorbars).
For most of the fitted values a bin width of 20 GeV/c$^2$ was used, except for the last bin where 100 GeV/c$^2$ were used to compensate for the fact that there are less 
events for higher invariant masses. From the expected errors one finds that
the parameter $R$ for the background can be estimated with a precision of about
$\sigma_{\rm R}$ = 0.08. This might not seem too good, but the effect of using a
 slightly wrong background distribution is not so large. 
To demonstrate this, a fit to the angular distribution of the angle $\theta$
was performed, where a wrong background distribution was subtracted
from the signal-plus-background distribution as obtained from the generator.
The parameter R$_{sub}$ of the subtracted distribution was changed to values 
higher and lower than the value of R$_{MC}$ of the generated distributions.  
In Table \ref{tab:wrongR}, the difference $\Delta$R = R$_{MC}$ - R$_{sub}$ and
the value of R$_{signal}$ obtained from the fit to the signal 
distribution produced 
by subtracting the wrong background distribution are shown.

\begin{table}[h]
\begin{tabular}{|c|c|c|c|c|c|}
\hline
$\Delta$R    & -0.2 & -0.1 & 0.0 & 0.1 & 0.2  \\ \hline
R$_{signal}$ & 0.747 & 0.758 & 0.770 & 0.782 & 0.796  \\ \hline
\end{tabular}
\caption{The measured Parameter $R$ for five different distributions that 
have been used to subtract the expected background distribution. 
$\Delta$R is the difference  between 
the value of $R$ from the background as produced by the Monte Carlo and the 
value of $R$ of the subtracted distribution.}
\label{tab:wrongR}
\end{table} 
The shift of the parameter is thus expected to be less than $\pm$0.01. 
Again, this is very small compared to the statistical error of $\Delta$R$_{stat}$ = 0.053. This error is not considered in the rest of the analysis. 
Furthermore, the effects will be even smaller when considering 
K-factors. Any K greater than 1 will give  better conditions to check
the background distributions. And, since the K-factor of the gluonic Higgs
production is higher than the K-factor of the main ZZ production process 
by quark antiquark pairs, the signal to background ratio will be even higher 
than predicted here.

\section{Results}

\begin{figure}
\epsfig{figure=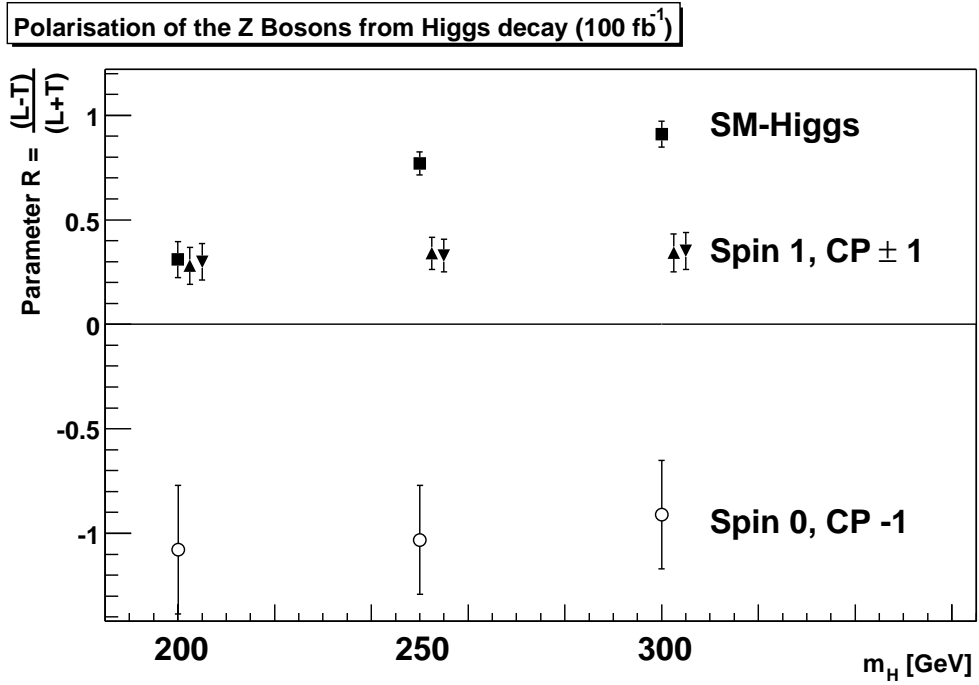,width=\textwidth}
\epsfig{figure=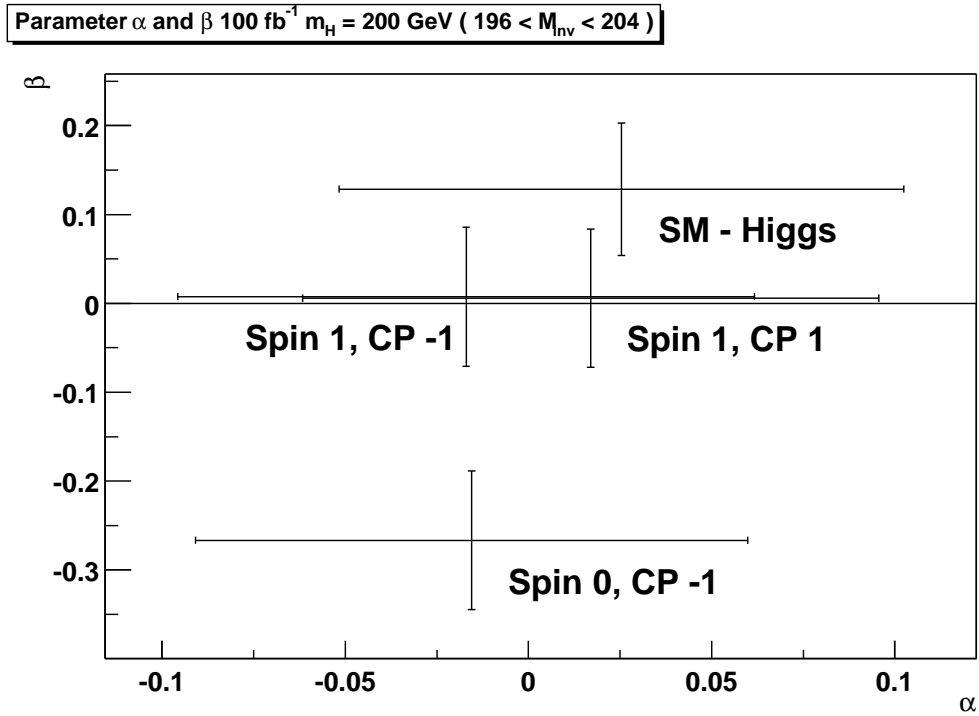,width=\textwidth}
\caption{The parameter $R$ for different Higgs massses (top) and $\alpha$ and $\beta$ (bottom) for $m_H$ =
  200 GeV/c$^2$ using 100 fb$^{-1}$. The error scales with the integrated luminosity as
  expected.}
\label{figpars}
\end{figure}

\begin{figure}
\epsfig{figure=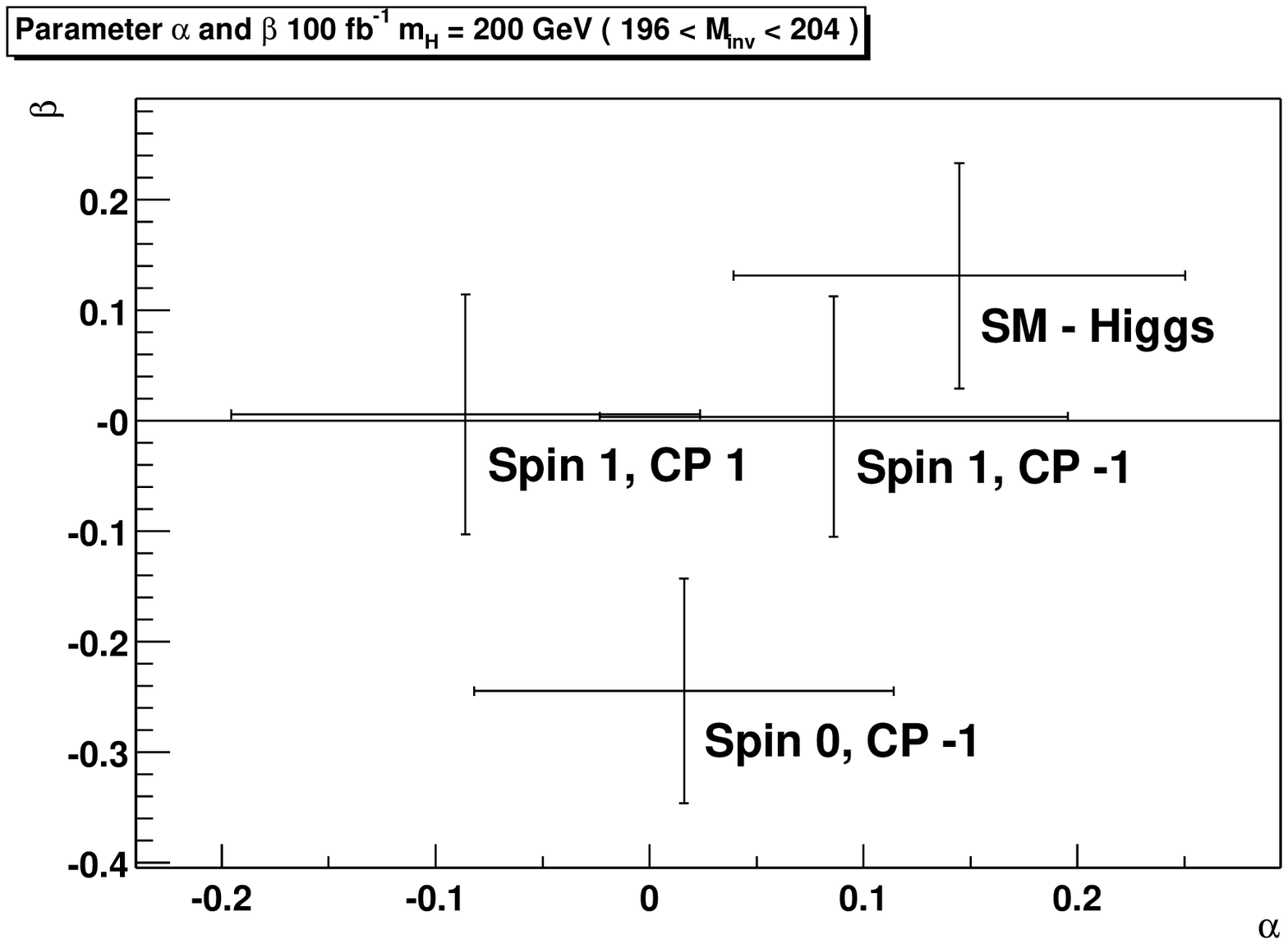,width=\textwidth}
\epsfig{figure=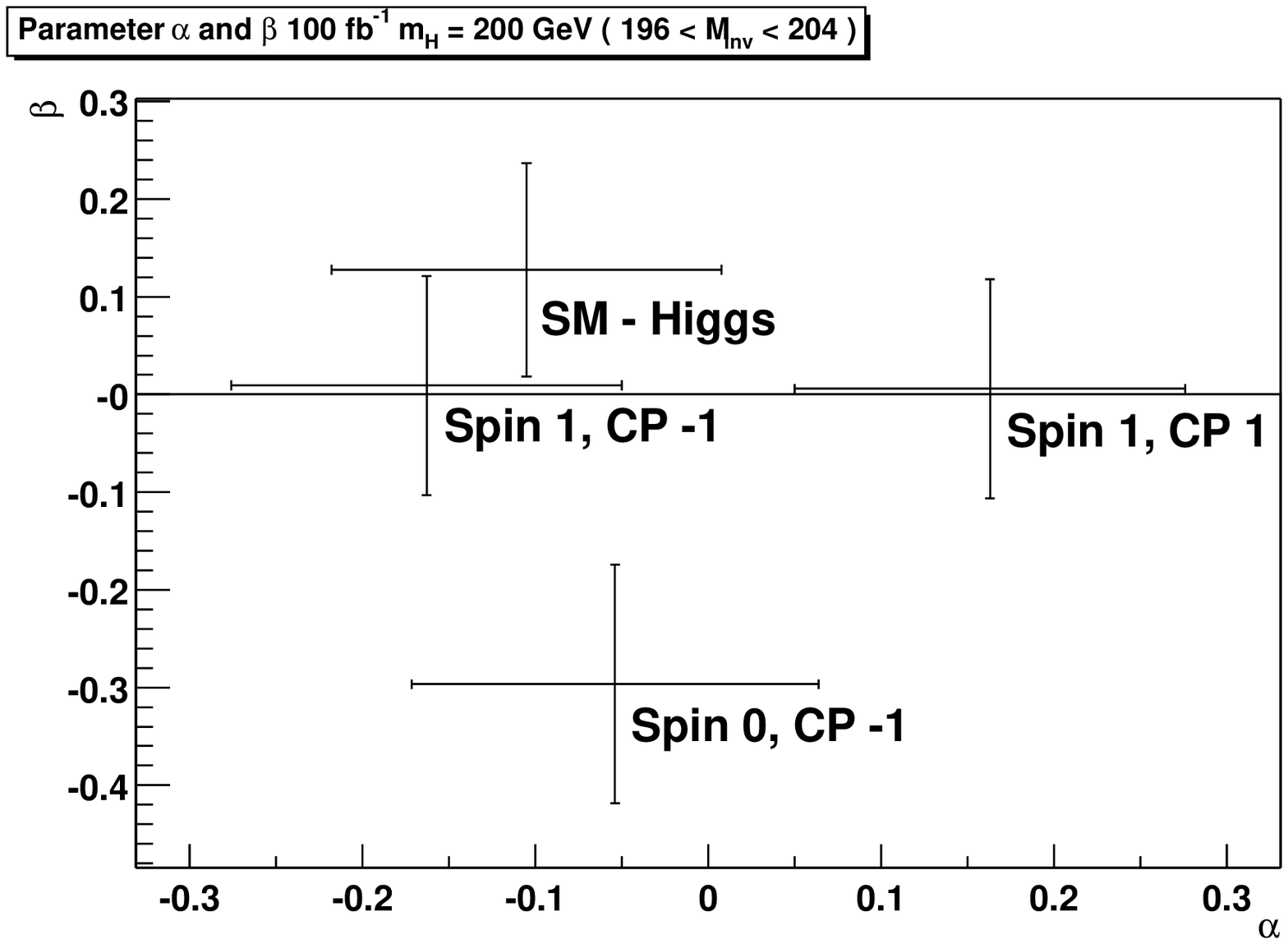,width=\textwidth}
\caption{The parameter $\alpha$ depends on the signs of the  $\cos(\theta)$ of
  the two Z bosons. The events where the signs are equal are used for the upper
  plot, those where the signs are different are used for the lower plot.}
\label{figsym}
\end{figure}

In Chapter 4, the exact results for the signal were given.
However, in practice one needs a procedure to separate signal from background,
which will lead to uncertainties in the distributions.
The expected errors have been calculated by generating a large number of events
and scaling the distributions to the expected number of events,
since the expected values of the parameters follow a Gaussian distribution.
The background was statistically subtracted after applying the same cuts to it 
as were applied to the signal. The error reflects the statistical error from
the number of the signal events, the statistical error from the number of 
background events subtracted and the error made by the estimation of the 
number of background events as described in 
Chapter 5. No error from a possibly different angular distribution of background
events has been taken into acount, but we have shown that the effect is small.
Then the parametrisations for $\phi$ and $\theta$ 
as described above were fitted to the distributions. Signal and background are summed 
over muons and electrons.\\

Figure \ref{figpars} (top) shows the expected values and errors for the parameter R, using 
an integrated luminosity of 100 fb$^{-1}$. It is clearly visible that for masses 
above 250 GeV/c$^2$ the measurement of this parameter allows the various hypotheses considered here 
for the spin and CP-state of the ``Higgs Boson'' to be unambiguously separated.

For a Higgs mass of 200 GeV/$c^2$ only the pseudoscalar is excluded.
Figure  \ref{figpars} (bottom) shows the expected values and
errors for $\alpha$ and $\beta$ for a 200 GeV/c$^2$ Higgs and an integrated luminosity of 
100 fb$^{-1}$.\\

The parameter
$\alpha$ can be used to distinguish between a spin 1 and the SM Higgs particle,
but its use is statistically limited.  The same applies to the parameter $\beta$.
Measuring $\beta$, which is zero for spin 1 and $>$ 0 in the SM case, can contribute only very 
little to the spin measurement even if $m_H$ is in the range where $\beta$, in the SM case, is close to 
its maximum value.
Nevertheless, $\beta$ can be useful to rule out a CP odd spin 0 particle.\\
 
The values of $\alpha$ get more widely separated when the correlation 
between the sign of $\cos(\theta)$ for the two Z Bosons and
$\phi$ is exploited. In Figure \ref{figsym}, we plot the parameters separately for
$sign(\cos \theta_1) = sign (\cos \theta_2)$ (F11 + F22 in Table \ref{distriint2}) 
and $sign(\cos\theta_1) = - sign(\cos \theta_2)$ (F12 + F21 in Table \ref{distriint2}). 
As can be seen, the difference in
$\alpha$ becomes bigger for $J=1$ and $\gamma_{CP}=+1$.
For higher masses $\alpha$ and $\beta$ of the SM Higgs approach 0; thus only
$\alpha$ can be used to measure the spin. But the measurement of R
compensates this.\\

\begin{figure}
\epsfig{figure=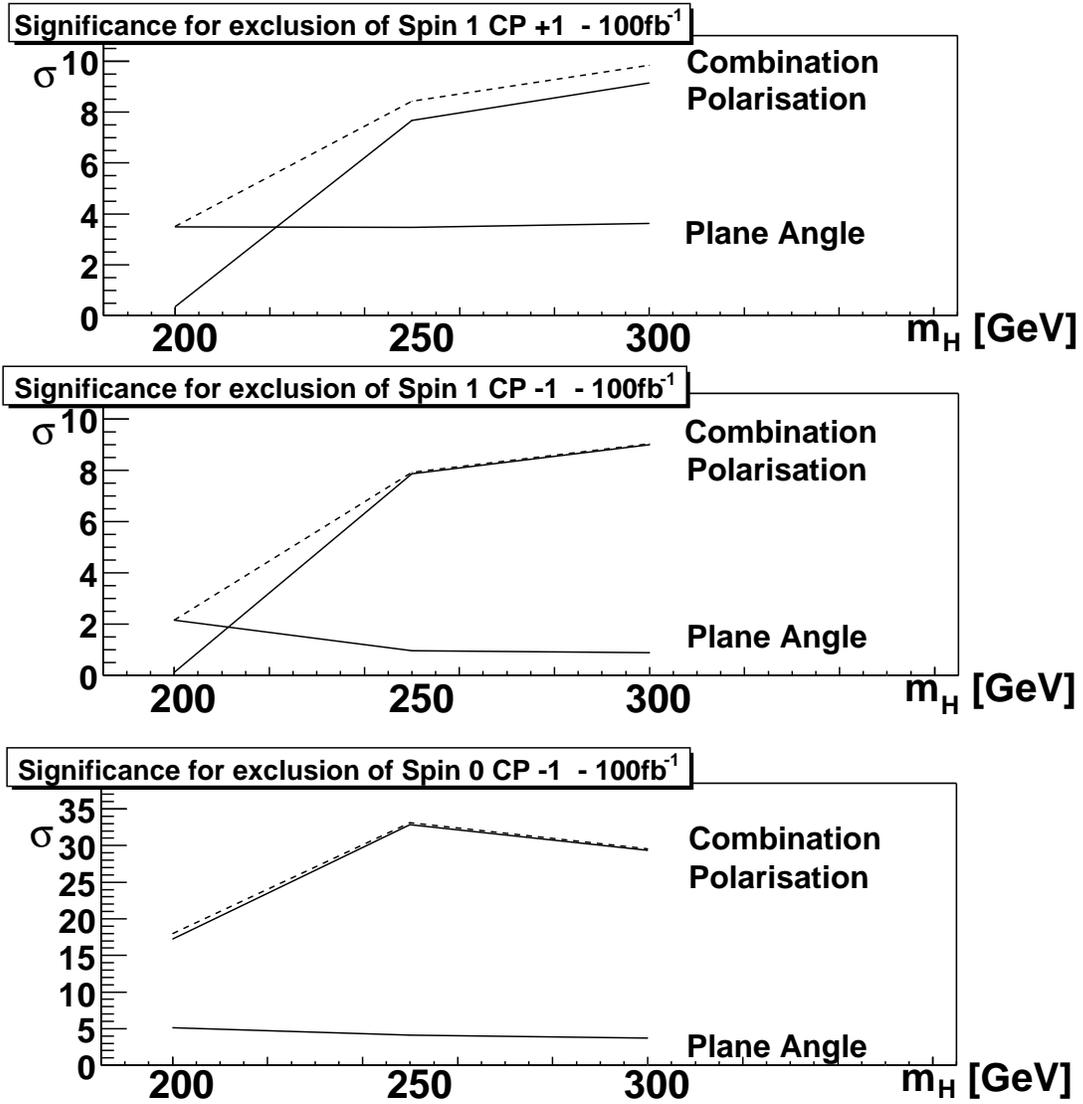,width=\textwidth}
\caption{
The overall significance for the exclusion of the non standard spin and CP-eigenvalue. 
The significance from the polar angle measurement and the 
decay-plane-correlation are plotted separately.}
\label{figsig}
\end{figure}

Figure \ref{figsig} shows the significance, i. e. the difference of the expected
values divided by  the expected error of the SM Higgs. We add up the
significance for $\alpha$ and $\beta$ exploiting the $\cos(\theta)$ - $\phi$
correlation and plot the sig\-ni\-fi\-cance from the polar angle measurement separately.
For higher Higgs masses the decay plane angle correlation contributes almost nothing, but the 
polarisation leads to a good measurement of the parameters spin and CP-eigenvalue.
For full luminosity (300 fb$^{-1}$) the significance  can simply be multiplied by $\sqrt{3}$. 
This is especially interesting for a Higgs mass of 200 GeV/c$^2$. The Spin 1, CP even
hypothesis can then be ruled out with a significance of 6.4$\sigma$, 
while for the Spin 1, CP odd case the significance is still only 3.9$\sigma$.\\

In conclusion, for Higgs masses larger than about 230\,GeV/c$^2$ a Spin 1 hypothesis can
be clearly ruled out already with 100 fb$^{-1}$. For m$_H$ around 200\,GeV/c$^2$ the 
distinction is less clear, and one will need the full 
integrated luminosity of the LHC. 
A  spin-CP hypothesis of 0- can be ruled out with less than 100 fb$^-1$ for the whole mass range 
above  and around 200 GeV/c$^2$.     


\section*{Acknowledgements} 
This work has been performed within the ATLAS Collaboration, and we thank collaboration 
members for helpful discussions. We have made use of the physics analysis framework 
and tools which are the result of collaboration-wide efforts. 
This work was supported by the {\it
DFG-Forschergruppe ``Quantenfeldtheorie, Computeralgebra und
Monte-Carlo-Simulation''} and
by the European Union under contract HPRN-CT-2000-00149.

\pagebreak

\appendix
\section{Formulae for  differential angular distributions}
The most general coupling of a (pseudo) scalar Higgs boson
to two on-shell Z-bosons is of the following form:

\begin{equation}
        {\cal L}_{scalar}=
           \mathbf{X} \delta_{\mu \nu}+
\mathbf{Y} k_{\mu} k_{\nu}/M_h^2 +i \mathbf{P} \epsilon_{\mu \nu p_{Z} q_{Z}}/M_h^2 
\label{scalar}
\end{equation}
Here the momentum of the first Z-boson is $p_Z^{\mu}$, 
that of the second Z-boson is $q_Z^{\nu}$.
The momentum of the Higgs boson is $k$ and  $\epsilon_{\mu\nu \rho\sigma}$ is the total antisymmetric tensor with $\epsilon_{1234} = i$.
Within the Standard Model one has $\mathbf{X}=1$, $\mathbf{Y}=\mathbf{P}=0$.
For a pure pseudoscalar particle one has $\mathbf{P} \not= 0, \mathbf{X}=\mathbf{Y}=0$.
If both $\mathbf{P}$ and one of the other interactions are present,
one cannot assign a definite parity to the Higgs boson.\\

The same formula for a (pseudo) vector with momentum
$k_{\rho}$ reads:
\begin{equation}
        {\cal L}_{vector}=
         \mathbf{X} (\delta_{\rho \mu} p_Z^{\nu}+\delta_{\rho \nu} q_Z^{\mu})
                  +\mathbf{P} (i \epsilon_{\mu \nu \rho p_Z}
   -i \epsilon_{\mu \nu \rho q_Z})
\label{vector}
\end{equation}
It is to be noted that the coupling to the vector field
actually contains only two parameters and is therefore simpler
than to the scalar.\\

In the following we give the angular dependence of the triple differential cross section for the case 
of a scalar or vector Higgs decaying into two on-shell Z bosons which 
subsequently decay into two lepton pairs. 
The meaning of the angles $\theta_1, \theta_2 $ and $\phi$ 
is explained in Figure \ref{figplanes}. $p$ is the absolute value of the momentum of the 
Z boson, $p^2 = (\frac{1}{2}M_h)^2 - M_Z^2$.
In the following we use the definitions $x = \frac{M_h}{M_Z}$ and $y =
\frac{p}{M_Z}$. $c_v$ and $c_a$ are the vector and axial vector couplings:
        $c_v = t_3 - 2q \sin(\theta_W)$,  $c_a = t_3$,
        where $t_3$ is the weak isospin, $q$ the charge of the fermion and 
        $\theta_W$ the Weinberg angle.
        For our case, the values of $c_v$ and $c_a$ are  $c_v$ = -0.0379 and $c_a$ = -0.5014.

\subsection{The general case}
\subsubsection* {Scalar Higgs}
\[
\begin{split}
&\frac{d\sigma}{d\phi d\cos \theta_1 d\cos \theta_2} \sim\,\\
& -8 {\mathbf X} 
{\mathbf Y} c_a^2 c_v^2 x^2 (x^2 -4 ) 
\cos \phi \sin \theta_1 \sin \theta_2 \\
 & - {\mathbf X} {\mathbf Y} (c_v^2 + c_a^2)^2 
 x^2 (x^2-4) (2 \cos \phi  \sin \theta_1
  \sin \theta_2 \cos \theta_1 \cos \theta_2  + 
( x^2-2)  \sin ^2\theta_1  \sin^2\theta_2  )  \\
 & + 16 {\mathbf X}{\mathbf P} c_a^2 c_v^2 x y (x^2-2) \sin \phi \sin \theta_1 \sin \theta_2 \\
 & + 4{\mathbf X}{\mathbf P}(c_v^2 + c_a^2)^2 x y \sin \phi (
 2 \cos \phi \sin ^2\theta_1  \sin ^2\theta_2  +(x^2- 2) 
 \sin \theta_1 \sin \theta_2\cos \theta_1 \cos \theta_2) \\
 & +16 {\mathbf X}^2 c_a^2 c_v^2 x^2 (2 \cos \theta_1 \cos \theta_2 +(x^2- 2) \cos \phi \sin \theta_1 \sin \theta_2 
   )  \\
 & + {\mathbf X}^2   (c_v^2 + c_a^2)^2  x^2   \{4 (1 + \cos ^2\theta_1  \cos ^2\theta_2  + 
 \cos ^2\phi \sin ^2\theta_1 \sin ^2\theta_2  \\
& +(x^2-2) \cos \phi  \sin \theta_1 \sin \theta_2 \cos \theta_1 \cos \theta_2) + 
   x^2 (x^2-4) \sin ^2\theta_1  \sin ^2\theta_2 \} \\
 & - 8 {\mathbf P}{\mathbf Y} c_a^2 c_v^2 xy (x^2-4) \sin \phi \sin \theta_1 \sin \theta_2   \\
 & - 2 {\mathbf P}{\mathbf Y}(c_v^2 + c_a^2)^2 x y  (x^2-4)   
\sin \phi \sin \theta_1 \sin \theta_2 \cos \theta_1 \cos \theta_2\\
 & + 1/4{\mathbf Y}^2  (c_v^2 + c_a^2)^2  
  x^2 (x^2-4)^2 \sin^2 \theta_1 \sin^2 \theta_2  \\
 & + 8 {\mathbf P}^2 c_a^2c_v^2  (x^2-4) \cos \theta_1 \cos \theta_2  \\
 & +{\mathbf P}^2(c_v^2 + c_a^2)^2 ( x^2-4)(1 +  \cos
  ^2\theta_1  \cos ^2\theta_2   - \cos ^2\phi  \sin^2\theta_1  \sin ^2\theta_2   )\\
\end{split}
\]

\subsubsection* {Vector Higgs}
\[
\begin{split}
&\frac{d\sigma}{d\phi d\cos \theta_1 d\cos \theta_2} \sim\,  \\
& -16 {\mathbf X} {\mathbf P}c_a^2 c_v^2 
    xy \sin \phi  \sin \theta_1  \sin \theta_2  \\
& + 4 {\mathbf X} {\mathbf P} (c_v^2+c_a^2)^2   x  y  
 \sin \phi  \sin \theta_1  \sin \theta_2 \cos \theta_1  \cos \theta_2\\
& + 4{\mathbf X}^2 c_a^2 c_v^2   x^2 \cos \phi  \sin \theta_1  \sin \theta_2  \\
& +  {\mathbf X}^2 (c_v^2+c_a^2)^2   x^2 (1 - \cos ^2\theta_1  \cos ^2\theta_2 
       - \cos \phi   \sin \theta_1  \sin \theta_2 \cos \theta_1  \cos \theta_2 )\\
& - 4  {\mathbf P}^2 c_a^2 c_v^2  (x^2-4)\cos \phi  \sin \theta_1  \sin \theta_2    \\
& + {\mathbf P}^2 (c_v^2+c_a^2)^2   (x^2-4)
     (1 - \cos ^2\theta_1  \cos ^2\theta_2  + 
  \cos \phi \sin \theta_1  \sin \theta_2  \cos \theta_1  \cos \theta_2  )
\end{split}
\]

\subsection{The special cases}
In this appendix we list the triple differential cross section for pure Higgs spin and CP states. 
In addition, we also give the differential cross sections, where some of the angular variables have 
been integrated over. F11, F12, F21, F22 refer to the different quadrants as defined in Chapter 4.
The spin 0, CP even part only contains  the pure SM contribution. 
\subsubsection* {Spin 0, CP even}
\[
\begin{split}
 \frac{d\sigma}{d\phi d\cos \theta_1 d\cos \theta_2} \sim\,&  
+ 16  c_a^2 c_v^2 
(2 \cos \theta_1 \cos \theta_2 +(x^2- 2) \cos \phi \sin \theta_1 \sin \theta_2 
  )  \\
  &+   (c_v^2 + c_a^2)^2 \{4 (1 + \cos ^2\theta_1  \cos ^2\theta_2  + \cos ^2\phi 
  \sin ^2\theta_1 \sin ^2\theta_2 \\
& +(x^2-2) \cos \phi  \sin \theta_1 \sin \theta_2 \cos \theta_1 \cos \theta_2) + 
 x^2(x^2-4)  \sin ^2\theta_1 \sin ^2\theta_2\}\\
 \frac{d\sigma}{d\cos \theta_1 d\cos \theta_2} \sim\,& 
 +32 c_a^2  c_v^2\cos \theta_1 \cos \theta_2  \\
&+  (c_v^2+c_a^2)^2 
   \{4 ( 1 +\cos^2\theta_1\cos^2\theta_2)+(x^4-4x^2+2)\sin^2\theta_1\sin^2\theta_2 \}\\ 
 \mbox{F11 = F22: }\frac{d\sigma}{d\phi} \sim\,&  c_a^2 c_v^2    (8 +  \pi^2 (x^2-2)\cos \phi)\\
& + 4/9 (c_v^2+c_a^2)^2 
 (  x^4  - 4 x^2 +10
    +( x^2-2)\cos \phi  + 4 \cos ^2\phi ) \\
 \mbox{F12 = F21: }\frac{d\sigma}{d\phi} \sim\, &
- c_a^2 c_v^2    (8 -  \pi^2 (x^2-2)\cos \phi)\\
& +4/9 (c_v^2+c_a^2)^2 
 (  x^4  - 4 x^2 +10
   - ( x^2-2) \cos \phi + 4 \cos ^2\phi ) \\
\end{split}
\]

\subsubsection* {Spin 0, CP odd}

\[
\begin{split}
 \frac{d\sigma}{d\phi d\cos \theta_1 d\cos \theta_2} \sim\, &
+8  c_a^2c_v^2   \cos \theta_1 \cos \theta_2  \\
&+(c_v^2 + c_a^2)^2 (1  + \cos^2\theta_1  \cos ^2\theta_2  
-   \cos ^2\phi  \sin ^2\theta_1  \sin ^2\theta_2  )\\\\
\frac{d\sigma}{d\cos \theta_1 d\cos \theta_2} \sim\,&
 +16 c_a^2 c_v^2\cos \theta_1 \cos \theta_2 + 
 (c_v^2+c_a^2)^2(1 + \cos ^2\theta_1) (1 + \cos ^2\theta_2) \\\\
\mbox{F11 = F22: }\frac{d\sigma}{d\phi} \sim\,&  c_a^2 c_v^2 + 1/9(c_v^2+c_a^2)^2  
  (5 -2 \cos ^2\phi) \\\\
\mbox{F12 = F21: }\frac{d\sigma}{d\phi} \sim\,& - c_a^2 c_v^2  + 1/9(c_v^2+c_a^2)^2 
 (5 - 2 \cos ^2\phi) 
\end{split}
\]
\subsubsection* {Spin 1, CP even}

\[
\begin{split}
\frac{d\sigma}{d\phi d\cos \theta_1 d\cos \theta_2} \sim\,&
+ 4 c_a^2 c_v^2 \cos \phi  \sin \theta_1  \sin \theta_2  \\
& +  (c_v^2+c_a^2)^2     (1 -\cos ^2\theta_1  \cos ^2\theta_2  - 
\cos \phi   \sin \theta_1  \sin \theta_2 \cos \theta_1  \cos \theta_2 )\\\\
\frac{d\sigma}{d\cos \theta_1 d\cos \theta_2} \sim\,&1 - \cos^2\theta_1 \cos^2\theta_2\\\\
\mbox{F11 = F22: }\frac{d\sigma}{d\phi} \sim\,&
+ c_a^2 c_v^2 \pi ^2\cos \phi +1/9(c_v^2+c_a^2)^2 (32 - 4\cos \phi)  
 \\\\
\mbox{F12 = F21: }\frac{d\sigma}{d\phi} \sim\,& 
   + c_a^2 c_v^2 \pi ^2\cos \phi  +1/9 (c_v^2+c_a^2)^2 (32 +4\cos \phi)\\
\end{split}
\]

\subsubsection* {Spin 1, CP odd}
\[
\begin{split}
\frac{d\sigma}{d\phi d\cos \theta_1 d\cos \theta_2} \sim\,& 
- 4   c_a^2 c_v^2  \cos \phi  \sin \theta_1  \sin \theta_2   \\
& + (c_v^2+c_a^2)^2   (1 - \cos ^2\theta_1  \cos ^2\theta_2  +
\cos \phi    \sin \theta_1  \sin \theta_2\cos \theta_1  \cos \theta_2
)\\\\
\frac{d\sigma}{d\cos \theta_1 d\cos \theta_2} \sim\,&1 - \cos^2\theta_1 \cos^2\theta_2\\\\
\mbox{F11 = F22: }\frac{d\sigma}{d\phi} \sim\,& 
  - c_a^2 c_v^2 \pi ^2\cos \phi+1/9 (c_v^2+c_a^2)^2 ) (32+4 \cos \phi)  
  \\\\
\mbox{F12 = F21: }\frac{d\sigma}{d\phi} \sim\,& 
  - c_a^2 c_v^2\pi ^2\cos \phi +1/9 (c_v^2+c_a^2)^2 ) (32-4\cos \phi)  
  \\\\
\end{split}
\]

\pagebreak


\begin{thebibliography}{200}

\bibitem{lephiggs} ALEPH, DELPHI, L3, OPAL and the Electroweak Working Group for Higgs Boson Searches, Phys. Lett. B565 (2003) 61.
\bibitem{zeuthen} The LEP Electroweak Working Group, LEP EWWG Home Page. URL: http://lepewwg.web.cern.ch/LEPEWWG (2003).
\bibitem{moenig} K. Moenig, Electroweak precision data and the Higgs mass: Workshop
                  summary, hep-ph/0308133 (2003). 
\bibitem{peskin} M. E. Peskin, Phys. Rev. D64 (2001) 093003.
\bibitem{piccinini} For a short review of the Higgs search status
at the LHC; F. Piccinini, contribution to ICHEP 2002, 
   hep-ph/0209377 (2002).
\bibitem{dominici} D.~Dominici, Riv. Nuovo Cim. {\bf 20}, 11 (1997).
\bibitem{kastening} B.~Kastening and J.~J.~van~der~Bij
Phys. Rev. {\bf D60}: 095003 (1999).
\bibitem{yang} C.~N.~Yang Phys. Rev. {\bf 77}, 242 (1950).
\bibitem{higgsreview} A recent review; C.~Quigg, 
Acta Phys. Polon. {\bf B30}, 2145 (1999).
\bibitem {distri1} A.~Abbasabadi and W.~W.~Repko, Nucl. Phys. {\bf B292},
(1987) 461 and Phys. Rev. {\bf D37}, (1988) 2668.
\bibitem{distri2}  M.~J.~Duncan, Phys.~Lett.~{\bf B179}, (1986) 393.
\bibitem{distri3} M.~J.~Duncan, G.~L.~Kane and W.~W.~Repko,
 Phys. Rev. Lett. {\bf 55},
 (1985) 773 and Nucl.~Phys.~{\bf B272}, (1986) 517.
\bibitem{distri4} J.~R.~Dell'Aquila and C.~A.~Nelson, Phys. Rev. {\bf D33},
 (1986) 80;\\
C.~A.~Nelson, Phys. Rev. {\bf D37}, (1988) 1220.
\bibitem{bij} T.~Matsuura and J.~J.~van~der~Bij, Z. Phys. {\bf C51},
259 (1991).
\bibitem{choi}   S. Y. Choi, D. J. Miller, M. M. Muhlleitner,
 P. M. Zerwas. Phys. Lett. B553 (2003) 61.	
\bibitem{cteq} H. L. Lai, J. Huston, S. Kuhlmann, F. Olness, J. Owens, D. Soper, W.K. Tung, H. Weerts. hep-ph/9606399. 1996
\bibitem{spira} A. Djouadi, J. Kalinowski, M.Spira. Comput. Phys. Commun. 108 (1998) 56.  
\bibitem{atlfast} {E. Richter-Was, D. Froidevaux, L. Pogglioli. ATL-PHYS-98-131. 1998}   
\bibitem{TDRV1}{Atlas Detector and Physics Performance. Technical Design Report. Volume 1. CERN 1999}
\bibitem{TDRV2}{Atlas Detector and Physics Performance. Technical Design Report. Volume 2. CERN 1999}

\end{thebibliography}
\end{document}